\documentclass[aps,prl,twocolumn]{revtex4}
\usepackage{amssymb}
\usepackage{graphicx}
\usepackage{amsmath}
\usepackage{bm}

\begin{document}

\title{Measuring Topological Number of a Chern-Insulator from Quench Dynamics}

\author{Ce Wang}
\affiliation{Institute for Advanced Study, Tsinghua University, Beijing, 100084, China}
\author{Pengfei Zhang}
\email{PengfeiZhang.physics@gmail.com}
\affiliation{Institute for Advanced Study, Tsinghua University, Beijing, 100084, China}
\author{Xin Chen}
\affiliation{Institute for Advanced Study, Tsinghua University, Beijing, 100084, China}
\author{Jinlong Yu}
\affiliation{Institute for Advanced Study, Tsinghua University, Beijing, 100084, China}
\author{Hui Zhai}
\email{hzhai@tsinghua.edu.cn}
\affiliation{Institute for Advanced Study, Tsinghua University, Beijing, 100084, China}
\date{\today}

\date{\today }

\begin{abstract}
In this letter we show how the topological number of a static Hamiltonian can be measured from a dynamical quench process. We focus on a two-band Chern insulator in two-dimension, for instance, the Haldane model, whose dynamical process can be described by a mapping from the $[k_x,k_y,t]$ space to the Bloch sphere, characterized by the Hopf invariant. Such a mapping has been constructed experimentally by measurements in cold atom systems. We show that, taking any two constant vectors on the Bloch sphere, their inverse images of this mapping are two trajectories in the $[k_x,k_y,t]$ space, and the linking number of these two trajectories exactly equals to the Chern number of the static Hamiltonian. Applying this result to a recent experiment from the Hamburg group, we show that the linking number of the trajectories of the phase vortices determines the phase boundary of the static Hamiltonian. 
\end{abstract}

\maketitle

Recently cold atom experiments have realized a number of topological models including the Hofstadter model \cite{Bloch_Hofstadter,Bloch_Hofstadter_Chern,Ketterle_Hofstadter}, the Haldane (and the Haldane-type) model \cite{Esslinger,USTC}, the Su-Schrieffer-Heeger model \cite{Bloch_SSH} and its Thouless charge pumping \cite{Bloch_pumping,Japan_pumping,Bloch_spin}. One major advantage of studying topological models in the context of cold atom systems, in comparison with its condensed matter counterpart, is that the experimental investigation of the dynamic processes can be more easily accessible. For example, considering non-interacting fermions initially in a topologically trivial insulator state of the initial Hamiltonian $\mathcal{H}^\text{i}$, we shall focus on a sudden quench to a final Hamiltonian $\mathcal{H}^\text{f}$, whose ground state is a topologically nontrivial insulator (e.g. a Chern insulator) at the same filling, and the question is whether the change of the topological number can be revealed from measuring the dynamical process after the quench. In fact, such a quench experiment has been performed recently in a Haldane-type model with cold atoms by the Hamburg group \cite{Sengstock}. Using a momentum resolved quantum state tomography method \cite{Sengstock2,Tomo1,Tomo2}, they can map out the evolution of the wave function as time evolves after the quench. 

At equilibrium, for a Chern insulator, it is known that the bulk Chern number, the number of edge states and the quantization value of the Hall conductance are equal, which is termed as `` the bulk-edge correspondence ". While for the non-equilibrium process after the quench, there is no such clear relations between them. First of all, because the time evolution after the quench is unitary, the Chern number of the quantum state does not change and does not reflect the topological number of the final Hamiltonian \cite{Rigol}. Nevertheless, the edge state gradually emerges \cite{Rigol, Cooper}. Second, without dephasing, the Hall response will not be well quantized for either a slow or a sudden quench \cite{Hu,Mueller}. While it is also found that the Hall conductance can become finite even after quenching to a topologically trivial final Hamiltonian \cite{Refael}. Therefore, it is desirable to know whether there is a way to rigorously map out the topology of the band structure of $\mathcal{H}^\text{f}$ through the quench dynamics. 

In this letter we present a scheme to extract a \textit{quantized} value from the dynamical process after the quench, and this quantized value is exactly the same as the topological Chern number of the final Hamiltonian $\mathcal{H}^\text{f}$. This scheme can be directly applied to analyze the recent experimental data from the Hamburg group (Ref. \cite{Sengstock}), as well as other similar systems (such as the ETH \cite{Esslinger} and the USTC experiments \cite{USTC}), to determine the topological phase diagram.  

\begin{figure}[tp]
\includegraphics[width=3.2 in]
{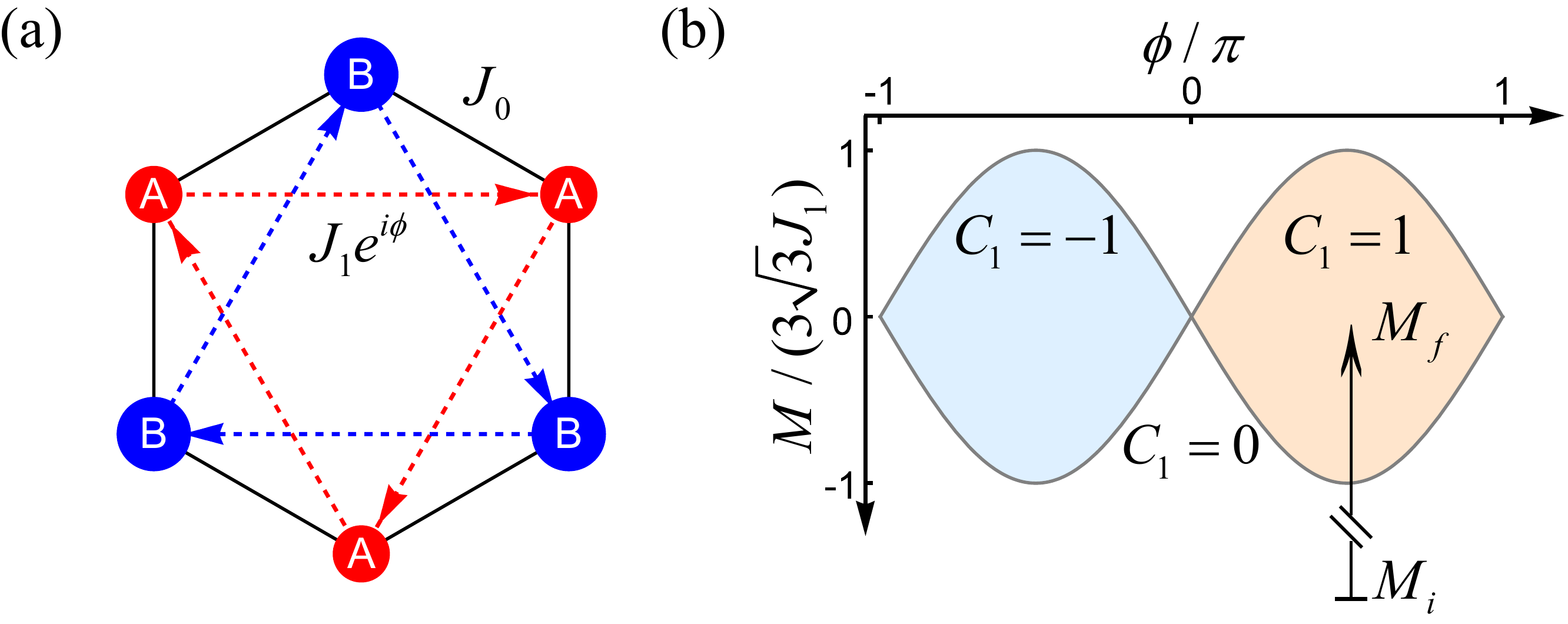}
\caption{(a) Schematic of hopping in the Haldane model in a honeycomb lattice; (b) The phase diagram of the Haldane model. The arrow indicates a quench from topologically trivial regime to a topologically nontrivial regime.   }
\label{Haldane_quench}
\end{figure}

\textit{Summary of the Scheme.} Before proceeding to details, let us briefly summarize our scheme as follows: 

Let us consider a general two-band tight-binding model in two-dimension, and at each momentum, the Hamiltonian can be written as
\begin{equation}
\mathcal{H}({\bf k})=\frac{1}{2}{\bf h}({\bf k})\cdot{{\bm \sigma}}, \label{matHk}
\end{equation}
where ${\bm \sigma}=(\sigma_x,\sigma_y,\sigma_z)$ is a vector of the Pauli matrices. Thus, the eigen-energies of the Hamiltonian are $\pm|{\bf h}({\bf k})|/2$, corresponding to the upper- and the lower-bands, respectively. We further consider at each ${\bf k}$, $|{\bf h}({\bf k})|$ is always non-zero, and the system is an insulator at half filling. The two-component wave function is denoted by $\zeta({\bf k})$. 

Here we consider the quench process that corresponds to a sudden change of ${\bf h}({\bf k})$ from a topologically trivial ${\bf h}^\text{i}({\bf k})$ to ${\bf h}^\text{f}({\bf k})$. The initial wave function $\zeta^\text{i}({\bf k})$ is taken as the lower-band eigenstate of the initial Hamiltonian. After the quench, the wave function will involve according to the final Hamiltonian as
\begin{equation}
\zeta({\bf k}, t)=\exp\left\{-\frac{i}{2}{\bf h}^\text{f}({\bf k})\cdot{\bm \sigma}t\right\}\zeta^\text{i}({\bf k}),  \label{evolution}
\end{equation}
and by introducing a Bloch vector 
\begin{equation}
{\bf n}=\zeta^\dag({\bf k}, t){\bm \sigma}\zeta({\bf k}, t), \label{nzeta}
\end{equation}
Eq. \ref{evolution} and Eq. \ref{nzeta} together define a mapping $f$ from $[k_x,k_y,t]$ to the Bloch sphere ${\bf n}$.

\textbf{The Scheme:} Taking any two constant vectors ${\bf n}_1$ and ${\bf n}_2$ on the Bloch sphere, their inverse images $f^{-1}({\bf n}_1)$ and $f^{-1}({\bf n}_2)$ are two trajectories in the $[k_x,k_y,t]$ space. The linking number of these two trajectories within the first Brillouin zone equals to the Chern number of the ground state for the final Hamiltonian at the same filling \cite{footnote_inverse}.

\begin{figure}[tp]
\includegraphics[width=3.4 in]
{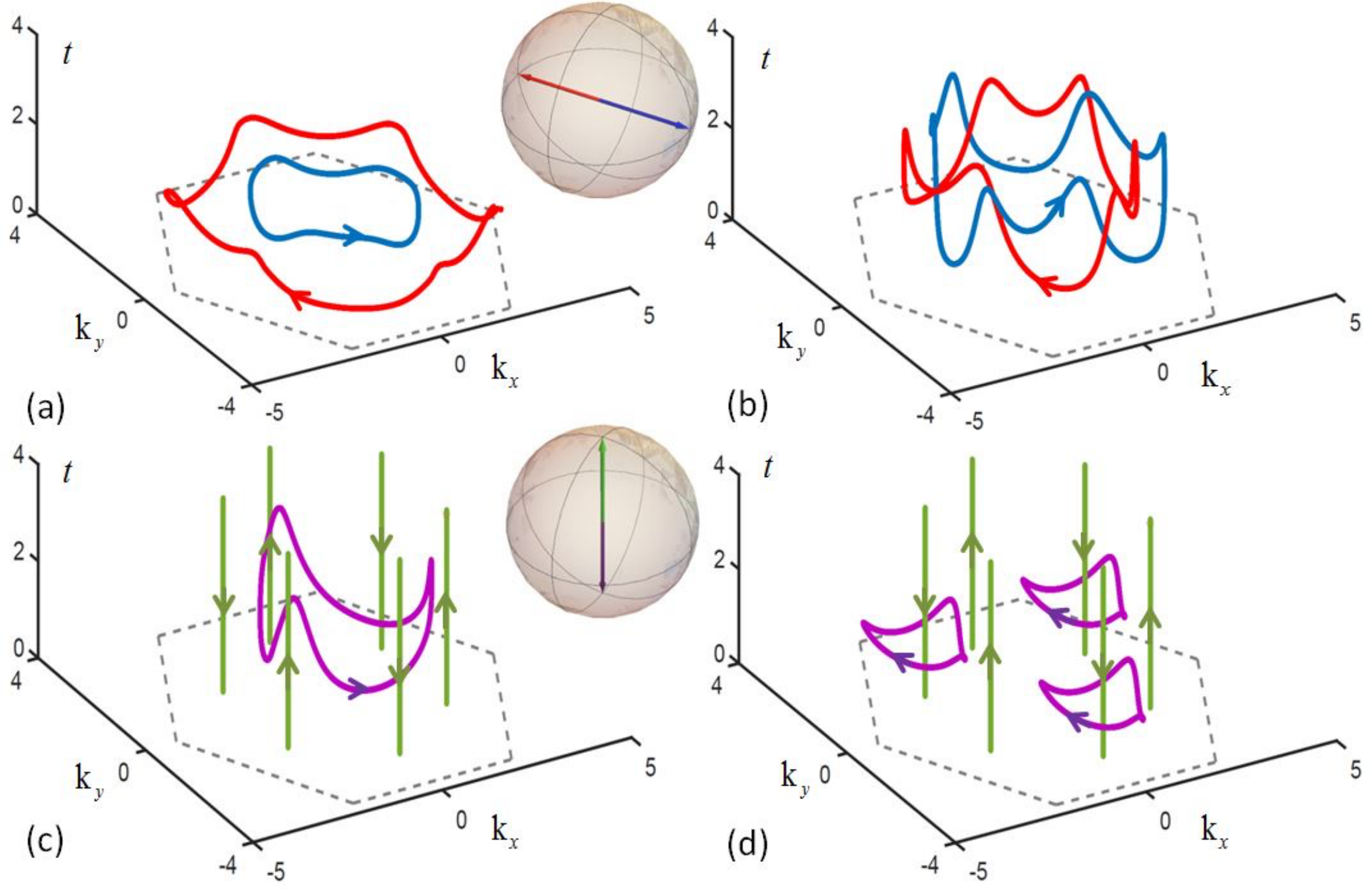}
\caption{(a-b) The inverse images of two vectors ${\bf n}$ and $-{\bf n}$ on the equator, when the Hamiltonian is quenched from ${\bf h}^\text{i}({\bf k})$ with $M=-\infty$ (topologically trivial regime) to ${\bf h}^\text{f}({\bf k})$ with $\phi=0.1$ and $M=1$ (topologically trivial regime)(a), and to ${\bf h}^\text{f}({\bf k})$ with $\phi=\pi/2$ and $M=0$ (topologically nontrivial regime)(b), respectively; (c-d) The inverse images of the north and the south poles, when the Hamiltonian is quenched from ${\bf h}^\text{i}({\bf k})$ with $M=-1$ and $\phi=\pi/2$ to ${\bf h}^\text{f}({\bf k})$ with $M=0.33\sqrt{3}$ and $\phi=\pi/2$ (topologically trivial regime) (c), and to ${\bf h}^\text{f}({\bf k})$ with  $M=0.27\sqrt{3}$ and $\phi=\pi/2$ (topologically nontrivial regime) (d). For all plots we have taken $J_0=1$ and $J_1=0.1$.  }
\label{Linking}
\end{figure}

\textit{Example to Illustrate the Scheme.} As a concrete example to illustrate our proposal, we consider the Haldane model in a honeycomb lattice [see Fig. \ref{Haldane_quench}(a)]. The particle annihilation operators at two sublattices of the honeycomb lattices are denoted by $\hat{a}_{{\bf r}_i}$ and $\hat{b}_{{\bf r}_i}$. The tight-binding model is written as 
\begin{align}
\hat{H}=&-J_0\sum\limits_{{\bf r}_i,j}\left(\hat{a}^\dag_{{\bf r}_i}\hat{b}_{{\bf r}_i+{\bf d}_j}+\text{h.c.}\right)+M\sum\limits_{{\bf r}_i}\left(\hat{a}^\dag_{{\bf r}_i}\hat{a}_{{\bf r}_i}-\hat{b}^\dag_{{\bf r}_i}\hat{b}_{{\bf r}_i}\right)\nonumber\\
&+J_1\sum\limits_{{\bf r}_i,j}\left(e^{-i\phi}\hat{a}^\dag_{{\bf r}_i}\hat{a}_{{\bf r}_i+{\bf a}_j}+e^{i\phi}\hat{b}^\dag_{{\bf r}_i}\hat{b}_{{\bf r}_i+{\bf a}_j}+\text{h.c.}\right), \label{Ham}
\end{align}
where ${\bf d}_{1,2}=(\pm\sqrt{3}/2,1/2)a_0$, ${\bf d}_3=(0,-1)a_0$ are the three vectors connecting the nearest neighboring sites; and ${\bf a}_{1,2}=(-\sqrt{3}/2,\pm 3/2)a_0$ and ${\bf a}_3=(\sqrt{3},0)a_0$ are the three vectors connecting the next nearest neighboring sites, with $a_0$ being the lattice spacing. The next nearest hopping has phase factor that is opposite between $A$ and $B$ sublattices. In the momentum space, Eq. \ref{Ham} becomes 
\begin{equation}
\hat{H}=\sum\limits_{{\bf k}}(\hat{a}^\dag_{{\bf k}},\hat{b}^\dag_{{\bf k}})\mathcal{H}({\bf k})\left(\begin{array}{c}\hat{ a}_{{\bf k}} \\ \hat{b}_{{\bf k}}\end{array}\right),
\end{equation}
and aside from a term proportional to the identity matrix, $\mathcal{H}({\bf k})$ takes the same form as Eq. \ref{matHk} with
\begin{align}
&h_x({\bf k})=-2J_0\sum_i\cos(\mathbf{k}\cdot\mathbf{d}_i), \label{hx}\\
&h_y({\bf k})=-2J_0\sum_i\sin(\mathbf{k}\cdot\mathbf{d}_i), \label{hy}\\
&h_z({\bf k})=2M+4J_1\sin\phi\sum_i\sin(\mathbf{k}\cdot\mathbf{a}_i). \label{hz}
\end{align}
The phase diagram of this Haldane model at half filling (with the lower-band filled) is shown in Fig. \ref{Haldane_quench}(b), where two topologically nontrivial regimes have the Chern numbers $+1$ and $-1$, respectively. Here we consider a sudden change of $M$ and $\phi$ starting from the topologically trivial regime, as indicated by the arrow in Fig. \ref{Haldane_quench}(b).

In Fig. \ref{Linking} we show two sets of examples. In Fig. \ref{Linking}(a-b), we consider the inverse image of two vectors ${\bf n}$ and $-{\bf n}$ on the equator. One can see that if $\mathcal{H}^\text{f}$ is in the topologically trivial regime, as shown in Fig. \ref{Linking}(a), $f^{-1}({\bf n})$ sets inside the trajectory of $f^{-1}(-{\bf n})$, and the linking number is zero; while if $\mathcal{H}^\text{f}$ is in the topologically nontrivial regime, as shown in Fig. \ref{Linking}(b), these two trajectories link three times. This is because, to avoid the discontinuity of the trajectory across the boundary of the first Brillouin zone, our plot spans the momentum regime including three replicas of the first Brillouin zone. Within the first Brillouin zone, the linking number is unity that equals to the Chern number of $\mathcal{H}^\text{f}$. Similarly, we consider the inverse images of the north and the south pole. As shown in Fig. \ref{Linking}(c-d), the inverse image of the north pole is a straight line in the $K$ and $K^\prime$ points. While the inverse image of the south pole does not enclose the $K$ or $K^\prime$ point if $\mathcal{H}^\text{f}$ is in the topologically trivial regime [Fig. \ref{Linking}(c)], giving rise to linking number zero; and it encloses three equivalent $K$ or $K^\prime$ point when $\mathcal{H}^\text{f}$ is topologically nontrivial [Fig. \ref{Linking}(d)], giving rise to linking number unity within the first Brillouin zone.    

Here we should put a remark of how to determine the sign of the linking number. First of all, we note that each trajectory actually has a direction, defining as the direction of ${\bf \mathcal{J}}$ with \cite{Zee}
\begin{equation} 
\mathcal{J}^\mu=\frac{1}{8\pi}\epsilon^{\mu\nu\lambda}{\bf n}\cdot(\partial_\nu{\bf n}\times\partial_\lambda{\bf n}), \label{Jmu}
\end{equation} 
with the indices taking $k_x$, $k_y$ and $t$. Thus, when two trajectories link and one trajectory crosses through a surface enclosed by another trajectory, if the direction of the trajectory is the same as the normal direction of the surface (determined by the right hand rule), we denote the linking number as $+1$ \cite{math}, otherwise we denote the linking number as $-1$. Hence, both cases shown in Fig. \ref{Linking}(b) and (d) have a linking number $+1$, consistent with the Chern number of $\mathcal{H}^\text{f}$ in the phase diagram Fig. \ref{Haldane_quench}(b). We have also checked that, if the Chern number of the $\mathcal{H}^\text{f}$ changes sign, the linking number defined in this way also changes sign.

\begin{figure}[tp]
\includegraphics[width=3.2 in]
{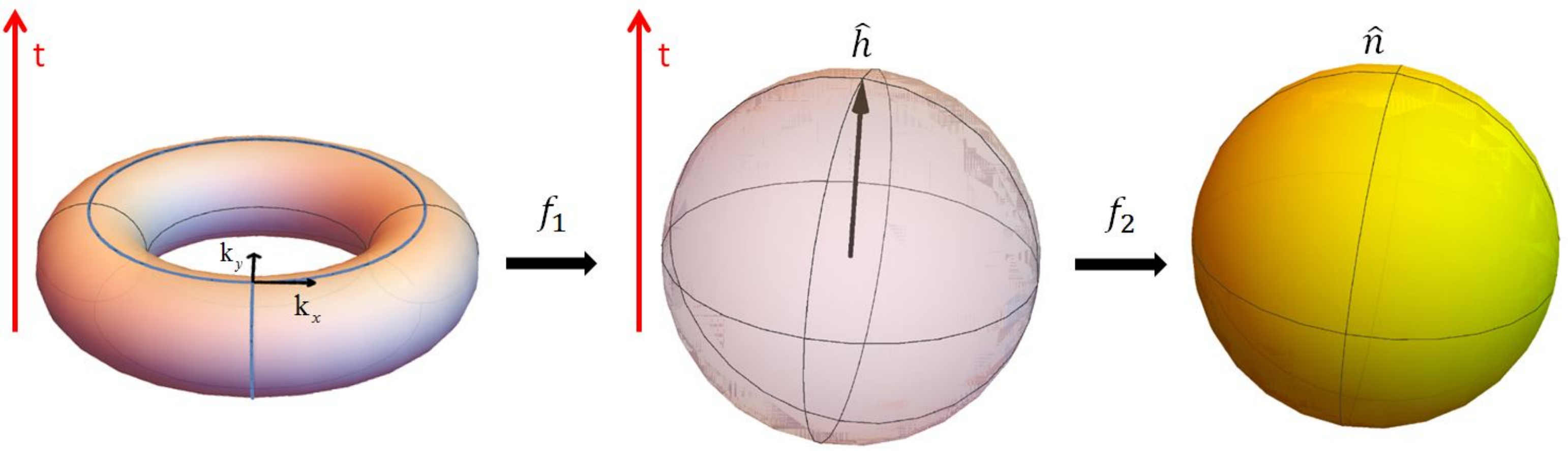}
\caption{Schematic of the mapping $f_1$ from $[k_x,k_y,t]$ ($[k_x,k_y]$ forms a torus) to $[{\bf h},t]$, and mapping $f_2$ from $[{\bf h},t]$ to ${\bf n}$ on a Bloch sphere.   }
\label{mapping}
\end{figure}

\textit{Mathematical Proof of the Results.} The general proof follows following three steps:

Step 1: From Eq. \ref{evolution} and Eq. \ref{nzeta}, one can see that at each ${\bf k}$, for $t=0$ and $t=2\pi/|{\bf h}^\text{f}({\bf k})|$, ${\bf n}$ always points to the north pole. Hence, they can be identified as one point. The mapping $f$ from $[k_x,k_y,t]$ to the Bloch sphere is thus a mapping from $T_3$ (three dimensional torus) to $S^2$, which is classified as $\mathcal{Z}$ for the situation considered here \cite{footnote}. The topological number is the Hopf invariant, and mathematically, it also equals to the linking number of two inverse images $f^{-1}$ \cite{math}. This number is invariant under continuous deformation of the mapping $f$. 

Step 2:  Now we decompose the mapping $f$ as $f_1\circ f_2$, as shown in Fig. \ref{mapping}. $f_1$ maps $[k_x,k_y]$ to ${\bf h}^\text{f}$ with the definition of ${\bf h}^\text{f}({\bf k})$ (e.g. Eq. \ref{hx}-\ref{hz}), which is classified by the Chern number $C$. $f_2$ maps $[{\bf h}^\text{f},t]$ to ${\bf n}$ with Eq. \ref{evolution} and Eq. \ref{nzeta}. 

If $f_1$ is a topologically trivial map, it can be continuously deformed into a mapping that all ${\bf k}$ points are mapped to the same vector. Thus, $[{\bf h}^\text{f}, t]$ forms an $S^1$. Then $f_2$ becomes a mapping from $S^1$ to $S^2$ that is always topologically trivial. Thus, $f$ is a topologically trivial map.  

In this step we focus on the case that $f_1$ is a topologically nontrivial map with non-zero Chern number $C$, we will show that one can construct a class of mapping $f$ such that the linking number of two inverse images of $f^{-1}$ equals $C$. For this construction, (i) we assume $|{\bf h}^\text{f}({\bf k})|=1$ for all momenta, thus, ${\bf h}^\text{f}$ itself is a Bloch sphere $S^2$ and the compact $[{\bf h}^\text{f},t]$ forms a $S^3$ ; (ii) we consider the initial wave function $\zeta^{i}({\bf k})=\zeta^0=\left(\begin{array}{c}1 \\0\end{array}\right)$ for all momenta, which corresponds to choosing ${\bf h}^i({\bf k})=(0,0,-1)$ or $M=-\infty$ in the example of the Haldane model; and (iii) we divide the first Brillouin zone into $|C|$ patches and we construct a mapping $f_1$ such that the boundary of each patch is mapped to the north pole of the Bloch sphere for ${\bf h}^\text{f}$, and therefore, different patches can be smoothly connected. 

We can stretch each patch into a compactified two-dimensional plane and parametrize this plane with polar coordinate $r$ and $\varphi$. For the convenience of computing the Hopf invariant, we can construct ${\bf h}^\text{f}$ as 
\begin{equation}
{\bf h}^\text{f}=(\cos\varphi\sin(f(r)),\sin\varphi\sin(f(r)),\cos(f(r))),
\end{equation}
where $\varphi$ changes between zero and $2\pi$, and $r$ changes between zero and infinity, with $f(0)=\pi$ and $f(\infty)=0$. 

Writing ${\cal J}^\mu=\epsilon^{\mu\nu\lambda}\partial_\nu B_\lambda$, the Hopf invariant is given by \cite{Zee}
\begin{equation}
H=\int rdrd\varphi dt\  B_\mu {\cal J}^\mu. \label{Hopf}
\end{equation}
With Eq. \ref{evolution}, \ref{nzeta}, \ref{Jmu} and this parametrization, it is straightforward to show that 
\begin{align}
{\cal J}^0&=\frac{1}{2\pi r}\sin^2\left(\frac{t}{2}\right)\sin(2f)f^\prime,   \label{J0}\\
{\cal J}^i&=-\frac{1}{4\pi}\sin f\left(\frac{r_i}{r^2}\sin t\sin f-2\epsilon_{ij}\frac{r_j}{r}\sin^2\left(\frac{t}{2}\right)f^\prime\right)    \label{Ji}
\end{align}
with $i$ being the two directions in the two-dimensional plane. Eq. \eqref{J0} and Eq. \ref{Ji} further give 
\begin{align}
&B_0=-\frac{1}{2\pi}\sin^2\left(\frac{t}{2}\right)\cos f,  \label{B0} \\
&B_i=-\frac{\epsilon_{ij}r_j}{2\pi r^2}\sin^2\left(\frac{t}{2}\right)\sin^2f. \label{Bi}
\end{align}
Substituting Eq. \ref{J0}-\ref{Bi} into Eq. \ref{Hopf}, one can obtain $H=1$ for this case. Thus, $f_2$ is a topologically nontrivial Hopf map. It is proved mathematically that the Hopf invariant equals to the linking number of the inverse images in $S^3$ of two different ${\bf n}$'s on $S^2$ \cite{math}, thus, the linking number of $f_2^{-1}$ equals unity. Furthermore, since there is a one-to-one correspondence between each patch in the first Brillouin zone and ${\bf h}^\text{f}$, thus, for the mapping constructed in this way, the linking number of the inverse mapping $f^{-1}$ equals $C$. Moreover, if the Chern number changes the sign, it corresponds to change $f(0)=0$ and $f(\infty)=\pi$ in the parametrization, and one obtains $H=-1$ for $f_2$. Consequently, the linking number also changes sign. 

Step 3: The three conditions used in the Step 2 can be released by continuously deforming the mapping, which does not change the linking number. For (i), as long as the system is gapped, i.e. $|{\bf h}^\text{f}|$ is finite everywhere, the length of $|{\bf h}^\text{f}|$ can always be continuously adjusted to $|{\bf h}^\text{f}|=1$. For (ii), as long as the initial wave function is in the same topologically nontrivial regime, it can be obtained by a continuous transformation of a uniform state as $\zeta^\text{i}({\bf k})=\mathcal{U}({\bf k})\zeta^0$, where $\mathcal{U}({\bf k})$ is smooth function of ${\bf k}$ and smoothly connects to the identity matrix. Hence, it does not change the linking number. Finally, continuously deforming $f_1$ also does not change the linking number, which releases the condition (iii). Therefore, we prove our results for a general situation.  

\textit{Application to Cold Atom Experiments.} Finally, we discuss applying our theory to recent cold atom experiments, such as the one from the Hamburg group \cite{Sengstock}. The initial state is prepared with all ${\bf n}$ pointing around the north pole, and they quench the system by turning on a periodic shaking which can induce gauge field and topological Haldane model \cite{Aoki,Zheng,Neupert,Esslinger,Sengstock,Sengstock2,Sengstock_gauge1,Sengstock_gauge2}. Using the method of momentum resolved quantum state tomography \cite{Sengstock2}, they are able to measure the wave function in the pseudo-spin bases as $\zeta({\bf k})=\left(\begin{array}{c}\sin(\theta_{{\bf k}}/2) \\ -\cos(\theta_{{\bf k}}/2)e^{i\varphi_{{\bf k}}}\end{array}\right)$. Two types vortices of the phase $\varphi_{{\bf k}}$ are found in the momentum space. The first type of the phase vortices are naturally located in the $K$ and $\Gamma$ points and their locations do not evolve with time, where ${\bf n}$ always points to the north pole; while the second type of phase vortices locate at certain momenta, at which ${\bf n}$ rotates to the south pole. The second type of vortices can be pair-wisely created and annihilated in the momentum space, tracing a trajectory in the $[k_x,k_y,t]$ space. 

In Ref. \cite{Sengstock}, they use the appearance of the second type of phase vortex as criterion to determine what they call a ``dynamical phase transition". And they find that the phase diagram for the ``dynamical phase transition" is much wider than the expected topological regime for the static Hamiltonian. With our results, the topological regime of the static Hamiltonian is determined by whether the trajectories of the second type of vortices wind around the trajectories of the first type of vortices, because these two trajectories are the inverse images of the south and the north poles, respectively. Thus, the linking number of these trajectories, as we shown in Fig. \ref{Linking}(c-d), determines the Chern number. For the Haldane model, it can be shown that, if the initial state uniformly points to the north pole for all momenta, once the second type vortices appear, their trajectories always wind around $K$ or $K^\prime$ points \cite{Yu}. However, if the initial state spreads a finite regime around the north pole, as in the case of the real experiment, there exists certain regime where the second type of vortices appear but their trajectory enclose neither $K$ nor $K^\prime$ \cite{Yu}. In these regime, the final Hamiltonians are still topological trivial ones. 

To summary, our result establishes a unique relation between quench dynamics and equilibrium property regarding the topological band structure. Our result can be apply to the Haldane model realized by the ETH \cite{Esslinger} and the Hamburg group \cite{Sengstock}, with measurements of the momentum resolved quantum state tomography measurement, and can also be applied to the Haldane-like model realized by the USTC group \cite{USTC}, where the real spin is used instead of the pseudo-spin, and momentum and spin resolved measurement will be sufficient. On the theory side, future generalizations include quench from topological nontrivial initial state, and quench in other classes of topological model, such as the time-reversal invariant $Z_2$ topological insulator. 

\textit{Acknowledgment.} This work is supported by MOST under Grant No. 2016YFA0301600 and NSFC Grant No. 11325418 and Tsinghua University Initiative Scientific Research Program.


\begin{thebibliography}{99}
\bibitem{footnote}
Considering a mapping from $T_3$ to $S^2$, if the mapping from each $T_2$ surface of $T_3$ to $S^2$ has zero Chern number, then the mapping from $T_3$ to $S^2$ is classified as $\mathcal{Z}$ \cite{LMDuan,Kennedy,math1}. In our case, mapping from $[k_x,k_y]$ to $S^2$ is a trivial mapping since the initial state has zero Chern number which does not change during the unitary evolution \cite{Rigol}. It can also be easily shown that, generically, for $[k_x,t]$, $[k_y,t]$ surfaces, ${\bf n}$ can not fully cover the Bloch sphere, and thus the mappings also have zero Chern number.
\end{thebibliography}
\end{document}